\documentclass[12pt]{iopart}

\usepackage{graphicx}
\usepackage{cite}
\usepackage{iopams}
\usepackage{bm}

\begin{document}

\title{Nuclear energy release from fragmentation}

\author{Cheng Li$^{1,2,}$\footnote{Email: imqmd@qq.com}, S R Souza$^{5}$, M B Tsang$^{1,2,3}$ and {Feng-Shou Zhang}$^{1,2,4}$\footnote{Corresponding author: fszhang@bnu.edu.cn}}

\address{$^{1}$The Key Laboratory of Beam Technology and Material Modification of Ministry of Education, College of Nuclear Science and Technology, Beijing Normal University, Beijing 100875, China\\
$^{2}$Beijing Radiation Center, Beijing 100875, China\\
$^{3}$National Superconducting Cyclotron Laboratory and Physics and Astronomy Department, Michigan State University,
East Lansing, Michigan 48824, USA\\
$^{4}$Center of Theoretical Nuclear Physics, National
Laboratory of Heavy Ion Accelerator of Lanzhou, Lanzhou 730000,
China\\
$^{5}$Instituto de F\'{\i}sica, Universidade Federal do Rio de Janeiro Cidade Universit\'{a}ria, Caixa Postal 68528, 21941-972, Rio de Janeiro, Brazil}
\begin{abstract}
Nuclear energy released by splitting Uranium and Thorium isotopes into two, three, four, up to eight fragments with nearly equal size are studied. We found that the energy released from equally splitting the $^{235,238}$U and $^{230,232}$Th nuclei into to three fragments is largest. The statistical multifragmentation model is employed to calculate the probability of different breakup channels for the excited nuclei. Weighing the the probability distributions of fragments multiplicity at different excitation energies for the $^{238}$U nucleus, we found that an excitation energy between 1.2 and 2 MeV/u is optimal for the $^{235}$U, $^{238}$U, $^{230}$Th and $^{232}$Th nuclei to release nuclear energy of about 0.7-0.75 MeV/u.
\end{abstract}

\maketitle

%\section{\label{int}Introduction}

Nuclear energy is one of the most efficient sources of energy based on the Einstein's mass-energy equivalence formula ($\Delta E=\Delta mc^2$). It originates in the strong force holding the protons and neutrons together. The release of nuclear energy happens when nucleons rearrange themselves to form more stable nuclei. In such nuclear reaction processes, a large amount of energy is released in the form of emitted kinetic energy of particles or fragments and electromagnetic radiation.

There are three ways to release nuclear energy, known as radioactive decay\cite{decay1,decay2,decay3,decay4}, fusion \cite{fus2,fus3,fus4,fus5,fus6,laws1} and fission \cite{fis1,fis3,fis4,fis5,fis6}. The radioactive decay is the spontaneous process which occurs in radioactive materials by which the nuclei of unstable parent nuclei, gradually break up and are transformed into more stable isotopes or into nuclei of a different type. The daughter nuclei, consequently lose energy by emitting radiation in the form of particles (include $\alpha$, $\beta$, photons, neutrons particle etc.) and/or electromagnetic waves. The continuous fusion reaction happens in the core of stars including our sun. However in order to harness it to produce power, the fusion process must be controlled. Nuclear energy from fusion is still far from commercially viable even though controllable fusion reaction has been studied for many years through the means of magnetic confinement \cite{mag1,mag2,mag3} and inertial confinement \cite{ine1,ine2,ine3,ine4}. The fission technology has been widely used to generate power by neutron-induced chain reactions. This fission process occurs when a heavy nucleus such as $^{235}$U and $^{239}$Pu absorbs a thermal neutron and the excited compound nucleus is excited to be near its fission barrier. At low excitation energy, quantum effects in the fission process are very significant to the dynamical evolution of the system. Nuclear structure effects, such as the shell corrections, directly influence the fission barrier height and the fission path. Most fission modes induced by thermal neutron are binary fissions in which one daughter nucleus has a mass of about 90 to 100 u and the remaining nucleus 130 to 140 u. Occasionally ternary fission and quaternary fission occur with relatively probability of 1/300 and 1/3000 per binary fission, respectively \cite{qian,qian2}. For each binary event of uranium nucleus, about 0.1 percent of the mass of the parent nucleus appears as the fission energy which is about 0.85 MeV per nucleon. For the nucleus with high excitation energies, the multifragmentation characterized by the multiple production of intermediate mass fragments is the typical decay mechanisms which is quite different from that at low excitation energies. It is not yet clear how much energy will be released in the multifragmentation process.

To describe the breakup process of excited nucleus, statistical models have been proposed, such as the statistical multifragmentation model (SMM), the statistical evaporation model (HIVAP) \cite{Hiv1,Hiv2} and the statistical model (GEMINI) \cite{gem1,gem2}, among others \cite{oth1,oth2,SMM5}. The SMM used to describe the breakup of excited nuclei in this work has been successfully applied to describe intermediate energy heavy-ion collisions \cite{SMM1} and spallation reactions \cite{SMM2}. The basic assumption is that, after the most violent stages of the reaction, the source at thermal equilibrium with mass number $A_0$, charge number $Z_0$, and excitation energy $E^*$ will expand and desintegrate almost simultaneously into a number of small and intermediate mass fragments. For a given excitation energy $E^*$, especially for highly excited nuclei, a large number of breakup channels are open. The different breakup channels are generated by Monte Carlo simulations. The probability of each breakup channel is normally formulated in terms of a statistical partition function ${W}_i$ which is related to the total energy and the Helmholtz free energy through the standard thermodynamic expression. The relative probability for a breakup channel can be written as $P=W_i/\sum{W}_i$. And the average value of a physical observable $O$ should be calculated through $<O>\equiv\sum O_iW_i/\sum{W_i}$. For more detailed description of the SMM see Refs. \cite{SMM1,SMM2,SMM3,SMM4,SMM6}.

%\section{\label{result}Results and discussion}

In the aftermath of the nuclear reaction, if the total mass of the final system is lower than the initial mass, the loss in mass which is called mass defect is converted into a large amount of energy given by ($\Delta E=\Delta mc^2$). The difference of total binding energy before and after the reactions is defined as:

\begin{equation}
\Delta E=\sum_{j (final)} E_b(A_j,Z_j)-\sum_{i (initial)} E_b(A_i,Z_i).
\end{equation}where $E_b(A_i,Z_i)$ denotes the binding energy of $i$-th nucleus.

The specific binding energy (binding energy per nucleon) reflects the interactions of the nucleons inside the nucleus, especially the short range nuclear force and the long range Coulomb forces. Fig. 1 shows the experimental specific binding energy for 2438 nuclei from the mass table AME2012 published in Refs. \cite{cpc}. $^{56}$Fe is the strongest bound nucleus and the nuclei with mass number around 56 have the largest specific binding energy values which is approximately 8.79 MeV/u. In general, for lighter nuclei with mass number less than 56, the nuclear force which is short range and attractive, has not reached saturation. Except for very light nuclei with increasing mass number, the average number of nucleons around one nucleon increases gradually, therefore, the specific binding energy also increases gradually. However, the Coulomb force, which is proportional to $Z^2$, is long range and repulsive and become more important with heavier nuclei. The specific binding energy for heavy nuclei decreases gradually with increasing of mass number.

In massive stars, the nuclear energy in a contracting core can be released through the neon, oxygen and silicon burning processes \cite{star1}. Complete exhaustion of the nuclear fuel in the stellar nucleosynthesis process happens when final production of the stable isotope $^{56}$Fe is formed. In a typical fission process, such as the reaction of n+$^{235}$U$\rightarrow ^{141}$Ba+$^{92}$Kr+3n, the total energy released is about 0.85 MeV/u which including 0.70 MeV/u and 0.15 MeV/u corresponds to the energy released in the fission process and decay process of prefragments $^{141}$Ba and $^{92}$Kr, respectively. Since the mass number of $^{141}$Ba and $^{92}$Kr are much larger than that $^{56}$Fe, the energy released in this fission process is incomplete. More nuclear energy can be released if such heavy nucleus can split into three or more fragments. Here we demonstrate the nuclear energy released for the three-fragments case. For simplicity, the decay process of unstable nuclei is not taken into account. We split a source with mass number $A$ into three fragments $A_1$, $A_2$ and $A_3$, and
\begin{equation}
A=A_1+A_2+A_3,
\end{equation}where, $A_i=0,1,2,3\cdots$ and each data point A is plotted in a Dalitz-type plot. For each set of fragments, the nuclear energy released by the source is calculated by Eq. (1) with mass table AME2012. Fig. 2(a) is a triangular Dalitz-type plot for $^{238}$U. In a triangular Dalitz-type plot one can represent each event with fragments' mass numbers $A_1$, $A_2$, and $A_3$ by a point placed at distances from three sides of the equilateral triangle. A point located at the corner, on the side and inside the triangle, corresponds to an event with one, two, and three fragments, respectively. In the central region of Dalitz-type plot, the energy emitted by $^{238}$U is 0.95 MeV/u as compared to 0.7$\sim$0.8 MeV/u released from splitting $^{238}$U into two equal size fragments. If a heavy source such as $^{238}$U splits equally into three fragments, the mass number of each fragments is about 80. To bring the fragments closer to $^{56}$Fe, it is necessary to split it into four fragments.

 Similar to the ternary fragment case, we represent four fragment events with fragments' mass numbers $A_1$, $A_2$, $A_3$ and $A_4$ by a point placed at distances from four surface of the tetrahedron. $A_4=0$ (Fig. 2(a)) denotes one of the four surface of the tetrahedron. Fig. 2(b) shows a slice of tetrahedron's Dalitz plot calculated with the mass table AME2012 with $A_4=60$ for $^{238}$U. From Fig. 2(b), one can see that 0.80 MeV/u is released if the $^{238}$U nucleus is splitted into four equal fragments. This is lower than 0.95 MeV/u obtained by splitting the nucleus into three equal fragments, despite the mass number of each of the four fragments be about 60 which is closer to 56. The fragments splitted from source $^{238}$U have larger neutron-proton ratio, the specific binding energy of these fragments is very small as the symmetry energy term plays an important role in nuclear binding energy.

Fig. 3 compares the average energy released per nucleon obtained with the mass table AME2012 for splitting $^{235}$U, $^{238}$U, $^{230}$Th and $^{232}$Th into two, three, four, five, six, seven, and eight fragments with nearly equal size. Nearly equal energy is released in the case of the two, four, and five fragments, which is about 0.8 MeV/u. The energy released from three fragments case is larger than in the other cases. Although splitting it into three fragments with nearly equal size is the most efficient way of releasing nuclear energy, the probability of ternary fission for the heavy nucleus at low excitation energies is very small. It is necessary to increase the excitation energy of the source to study the nuclear energy released in the multifragmentation process. However the multifragmentation process for highly excited nuclei is very complex, as a large number of decay channels are simultaneously open \cite{souza}. We employ statistical multifragmentation model (SMM) which is a classical statistical model to calculate the probability of different decay channels for the excited nucleus.

 Fig. 4 shows the probability as a function of the fragment multiplicity for$^{238}$U sources with excitation energy $E^*=$1, 2, 3 and 4 MeV/u. We select events whose multiplicity values lie between 2-10. It is seen that many breakup channel are simultaneously open for a given excitation energy. For $E^*=1$ MeV/u, the probability are approximately 0.6 for binary events and 0.4 for ternary events. As the excitation energy increases, decay channels with larger multiplicities are opened gradually and the position of the maximum probability is moves to higher multiplicity values.

In the framework of SMM, the average energy released is obtained by averaging this quantity in each fragmentaion mode $<\triangle{E}>\equiv\sum \Delta E_i W_i/\sum W_i$, where $\triangle{E_i}$ denotes the nuclear energy release in the $i$-$th$ event and $W_i$ is corresponding statistical weight. The average nuclear energy released per nucleon for the $^{235}$U, $^{238}$U, $^{230}$Th, and $^{232}$Th at different excitation energies are shown in Fig. 5. As is well known, for a heavy nuclei with low excitation energy, one observes the emission of light particles plus an evaporation residue in the deexcitation process. Fission occurs usually when the excited energy is larger than the fission barrier of the nuclei. Multifragmentation often happens in highly excited nuclei. From Fig. 5 one can see that at $E^*=$1 MeV/u, the average nuclear energy release rises up to 0.7 MeV/u, as one varies the source nucleus, due to the opening of the binary breakup channel. Between 1.2 and 2 MeV/u, the probability of binary, ternary and quaternary breakup channel accounts for the largest proportion of the emitted fragments, and peaks around 0.7$\sim$0.75 MeV/u. However as the excitation energy surpasses 2 MeV/u, the breakup channels with higher multiplicities dominates the partition function. The average energy released per nucleon gradually decreases with the increasing of the excitation energy.

%Fig. 7 shows that the probability as a function of fragments multiplicity for the $^{238}$Th, $^{238}$U and $^{238}$Np with excitation energy $E^*=1, 2, 3$ and 4 $MeV/u$. The black, red, blue and green line denote the excitation energy $E^*=1, 2, 3$ and 4 $MeV/u$, respectively. One can see from that the probability of fragments multiplicity for $^{238}$Th is higher than that the $^{238}$U and $^{238}$Np in the right of peaks. It is due to the neutron-rich nuclei emit more neutrons in the thermodynamic expansion of high excitation nuclei.
%\begin{figure}[h]
%\includegraphics[width=8cm,angle=0]{fig7.eps}
%\caption{(color online).The probability as a function of fragments multiplicity for the $^{238}$Th, $^{238}$U and $^{238}$Np with excitation energy $E^*=1, 2, 3$ and 4 $MeV/u$. }
%\end{figure}

In summary, the nuclear energy released has been calculated by splitting $^{235}$U, $^{238}$U, $^{230}$Th and $^{232}$Th sources into two, three, four, five, six, seven, and eight fragments with nearly equal size. The calculation results demonstrate that fragmentation into three nearly equal fragments is the most efficient mechanism to release the nuclear energy. We found that 0.95 MeV per nucleon is released in this case. The statistical multifragmentation model has been employed to simulate the breakup process of excitation nucleus. Taking into account the probability distributions of fragments multiplicity at different excitation energies we find that an excitation energy between 1.2 and 2 MeV/u is optimal for the $^{235}$U, $^{238}$U, $^{230}$Th and $^{232}$Th nuclei to release nuclear energy.

\section*{Acknowledgments}

We wish to thank J. Randrup for fruitful discussion. This work was supported by the National Natural Science
Foundation of China under Grants No. 11025524 and No. 11161130520
and the National Basic Research Program of China under Grant No.
2010CB832903.

\newpage

\begin{figure}
\begin{center}
\includegraphics[width=12cm,angle=0]{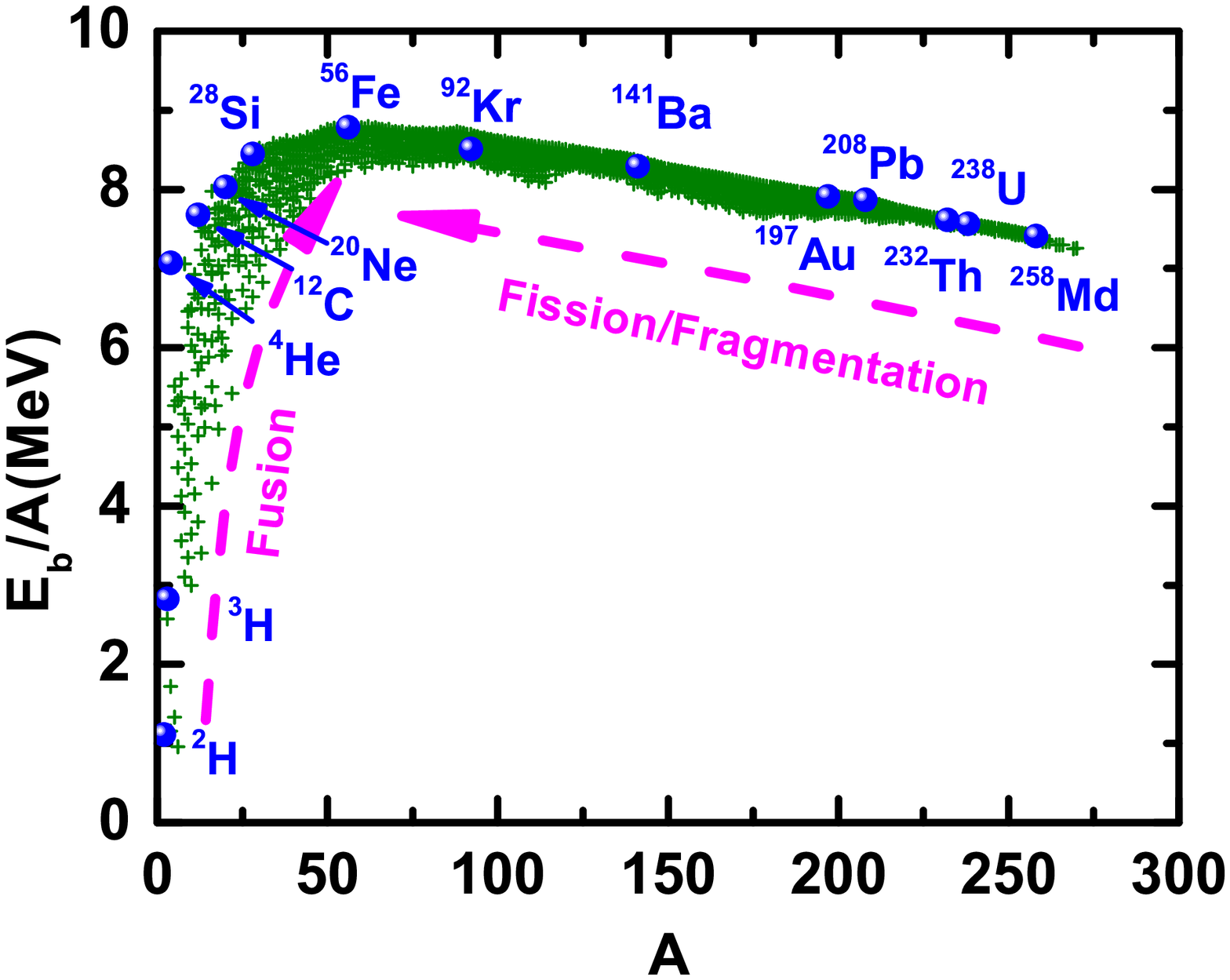}
\caption{\label{dri} Empirical specific binding energy as a function of mass number A for 2438 nuclei from the new mass table AME2012 \cite{cpc}. }
\end{center}
\end{figure}

\begin{figure}
\begin{center}
\includegraphics[width=12cm,angle=0]{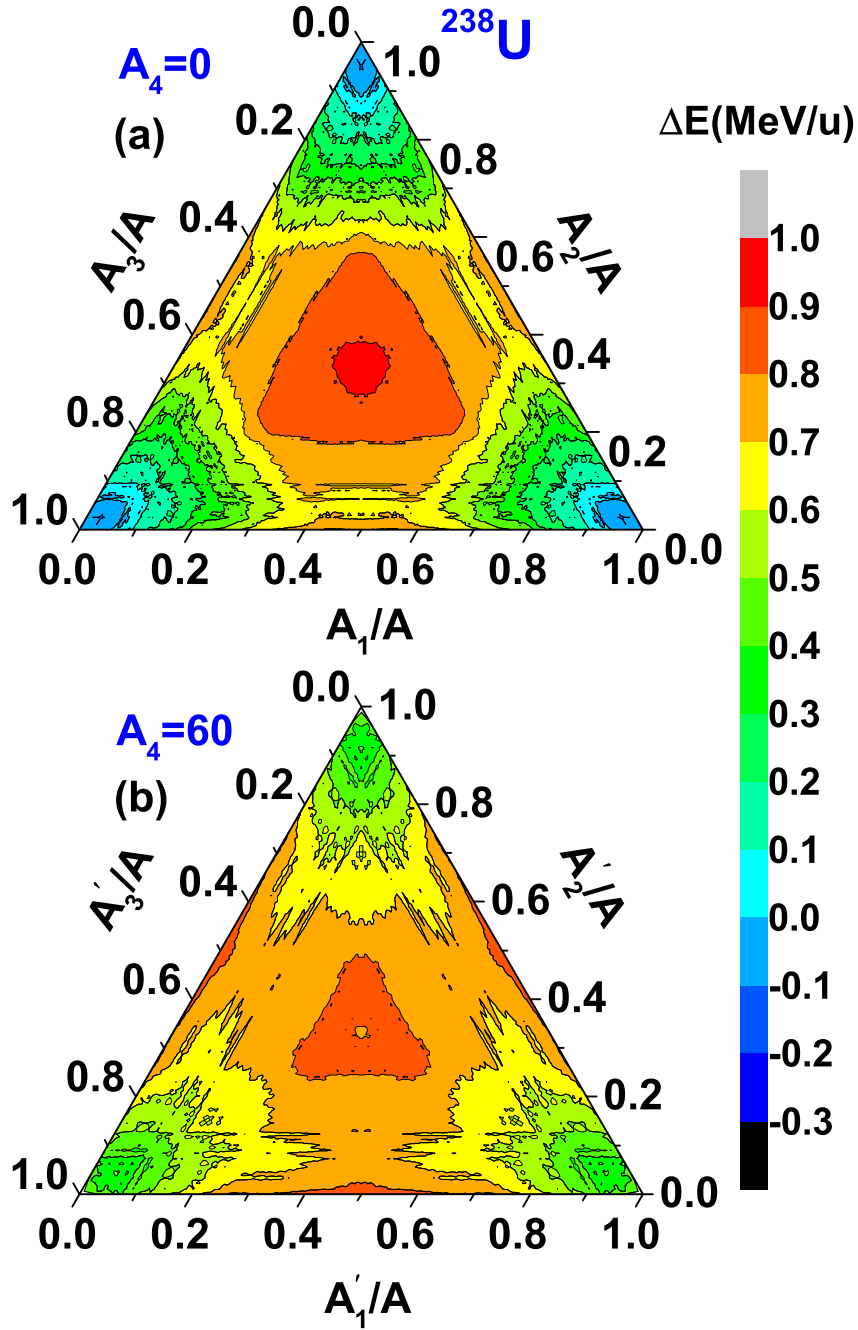}
\caption{\label{dis} (a) The triangular Dalitz-type plot for the nuclear energy released obtained with empirial values by splitting $^{238}$U into three fragments. The mass numbers of the three fragments and the source are denoted by $A_i$ ($i=1, 2, 3$) and $A$, respectively. (b) The slice of tetrahedron's Dalitz plot for $^{238}$U with $A_4=60$. The $A_i^\prime$ denotes the distances between one point and one of three side on slice and satisfy relation $A_i^\prime=3\sqrt{2}/4A_i$ ($i=1,2,3$) which is obtained by geometric relationships. }
\end{center}
\end{figure}

\begin{figure}
\begin{center}
\includegraphics[width=12cm,angle=0]{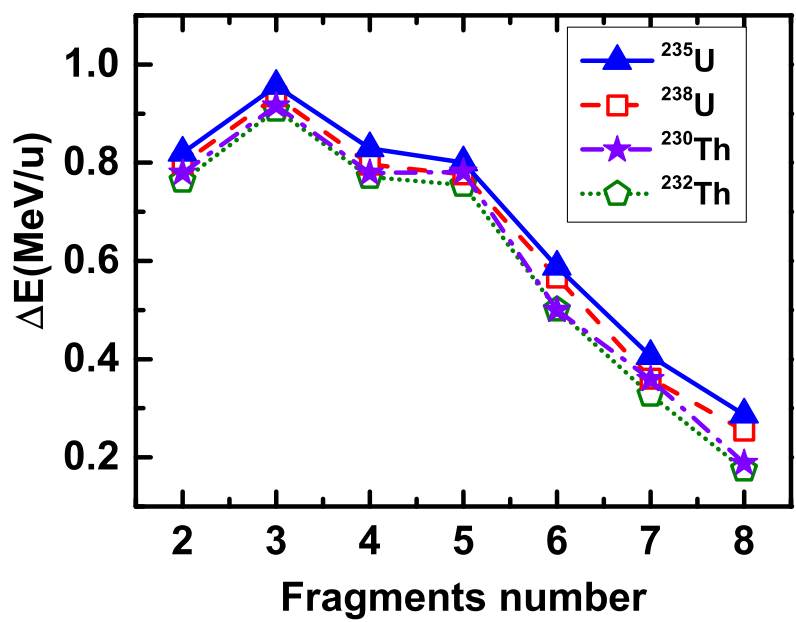}
\caption{\label{cross} Average energy released per nucleon for $^{235}$U, $^{238}$U, $^{230}$Th and $^{232}$Th split to 2, 3, 4, 5, 6, 7, and 8 fragments with nearly equal size. }
\end{center}
\end{figure}

\begin{figure}
\begin{center}
\includegraphics[width=12cm,angle=0]{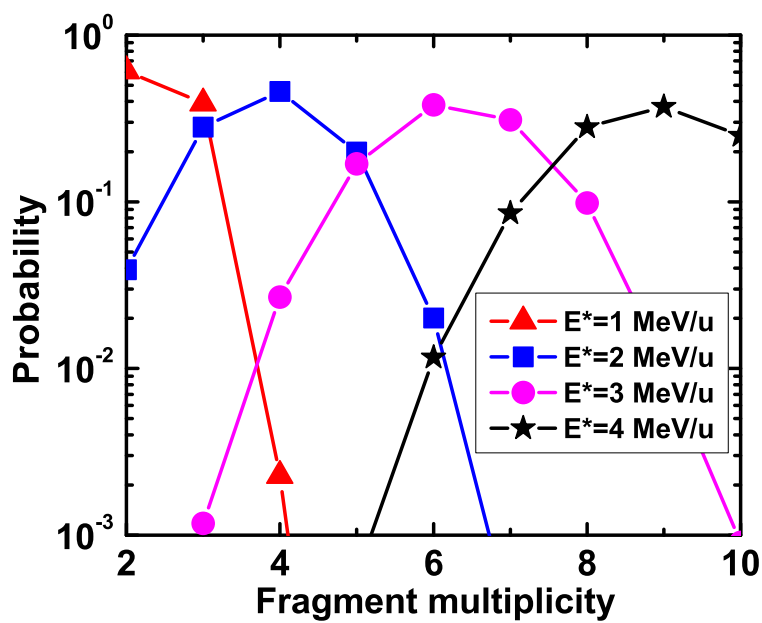}
\caption{\label{proton} The normalized multiplicity distributiom as a function of fragment multiplicity obtained with SMM for $E^*=$1, 2, 3 and 4 MeV/u.}
\end{center}
\end{figure}
\begin{figure}
\begin{center}
\includegraphics[width=12cm,angle=0]{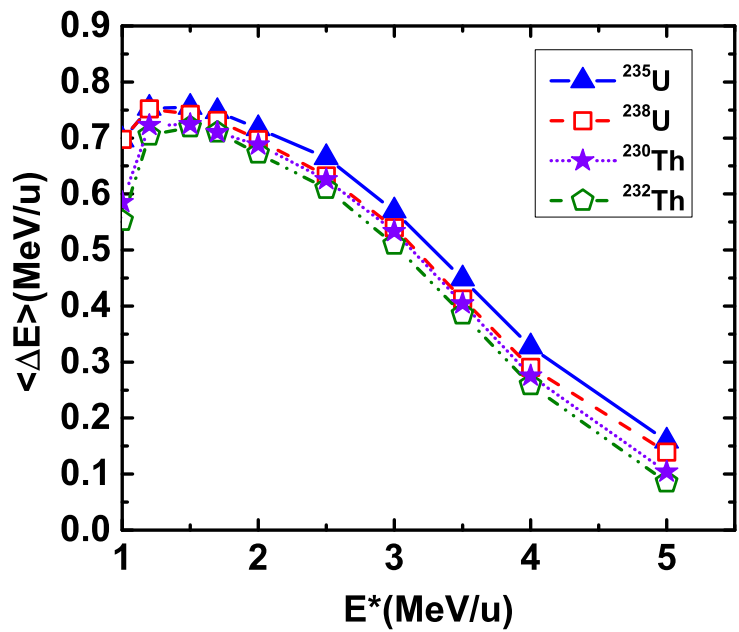}
\caption{\label{neutron} Average energy release as a function of the
excitation energy $E^*=1, 2, 3$ and 4 MeV/u for the $^{232,238}$Th and $^{235,238}$U sources.}
\end{center}
\end{figure}

\end{document}